\documentclass{article}

\usepackage{graphicx}
\usepackage[dvipsnames]{xcolor}
\usepackage{amsmath}
\usepackage{amssymb} 
\usepackage{geometry} 
\usepackage[backend=biber, style=nature , doi=true]{biblatex}
\begin{document}

\begin{center}
  \textbf{\LARGE{Super-resolution imaging with patchy microspheres}}
  
  \vspace{0.5cm}
  
\large{Qingqing Shang,$^1$ Fen Tang,$^2$ Lingya Yu,$^3$ Hamid Oubaha,$^4$ Darwin Caina,$^{4,5}$ Sorin Melinte,$^4$ Chao Zuo,$^6$ Zengbo Wang,$^{3,7}$ and Ran Ye$^{2,6,8}$}

\end{center}

\noindent $^1$Key Laboratory for Opto-Electronic Technology of Jiangsu Province, Nanjing Normal University, Nanjing, 21033, China

\vspace{0.2cm}

\noindent $^2$School of Artificial Intelligence, Nanjing Normal University, Nanjing, 21033, China

\vspace{0.2cm}

\noindent $^3$School of Computer Science and Electronic Engineering, Bangor University, Bangor, LL57 1UT, UK

\vspace{0.2cm}

\noindent $^4$Institute of Information and Communication Technologies, Electronics and Applied Mathematics, Universite\'e catholique de Louvain, Louvain-la-Neuve, Belgium, 1348

\vspace{0.2cm}

\noindent $^5$Faculty of Engineering, Physical Sciences and Mathematics, Universidad Central del Ecuador, Quito, 170402 Ecuador

\vspace{0.2cm}

\noindent $^6$Smart Computational Imaging Laboratory (SCILab), School of Electronic and Optical Engineering, Nanjing University of Science and Technology, Nanjing, 210094, China

\vspace{0.2cm}

\noindent $^7$z.wang@bangor.ac.uk

\vspace{0.2cm}

\noindent $^8$ran.ye@njnu.edu.cn

\vspace{1cm}
\noindent \textbf{Abstract:} The diffraction limit is a fundamental barrier in optical microscopy, which restricts the smallest resolvable feature size of a microscopic system. Microsphere-based microscopy has proven to be a promosing tool for challenging the diffraction limit. Nevertheless, the microspheres have a low imaging contrast in the air, which hinders the application of this technique. In this Letter, we demonstrate that this challenge can be effectively overcome by using partially Ag-plated microspheres. The deposited Ag film acts as an aperture stop that blocks a portion of the incident beam, forming a photonic hook with oblique near-field illumination. Such a photonic hook significantly enhanced imaging contrast, as experimentally verified by imaging Blu-ray disc surface and silica particle arrays.

\section*{Introduction}

Optical microscopes (OMs) are one of the most important tools for scientific research. Due to the Abbe diffraction limit, conventional OMs cannot resolve two objects closer than 0.5$\lambda$/NA, where $\lambda$ is the incident wavelength and NA the numerical aperture of the microscope. Therefore, an OM equipped with a near-unity NA objective and a white light source ($\lambda$ $\sim$ 550-600 nm) has a resolution limit of 300 nm. Within this context, many different methods have been proposed to overcome this limitation. In 2011, Wang et al. demonstrated the 50 nm-resolution of dielectric microspheres with white light illumination~\cite{11WangZ}. Since then, super-resolution imaging using dielectric microspheres has attracted considerable scientific interest. Microsphere-assisted imaging has the advantages of simple operation, no fluorescent labelling and good compatibility with commercial OMs. To obtain high-quality images, various parameters affecting the imaging performance of microspheres have been studied, such as illumination conditions~\cite{19PerrinS}, microsphere diameters~\cite{16GuoM}, immersion mode~\cite{12DarafshehA, 18WangF, 18ZhouY} and immersion materials~\cite{13LeeS, 11HaoX, 15DarafshehA}.

Currently, most microsphere-assisted imaging methods use high-refractive-index microspheres in a liquid environment~\cite{17DarafshehA, 13YeR, 17HouB, 14YangH, 17HuszkaG, 13LiL}. Nevertheless, the imaging process is affected by the shape of the air-liquid interface as well as the refractive index distribution inside the liquid. Moreover, samples may be contaminated or damage in liquid. Only a few research has been done to improve the imaging performance of microspheres in air, such as improving illumination conditions~\cite{19PerrinS}, optimizing the diameter and the refractive index of microspheres~\cite{13LeeS2}, using plano-convex-microsphere (PCM) lens design~\cite{20YanB, 20YanB2} and using a microsphere lens group~\cite{20LuoH}.

This Letter presents the performance of super-resolution imaging in the air using patchy microspheres. To the best of our knowledge, this is the first study showing that patchy microspheres are suitable for super-resolution imaging. The patchy particles can generate a curved photonic jet, i.e. photonic hook, due to the structural asymmetry~\cite{18YueL, 19MininIV, 19GuG}, which is shown to be useful in boosting imaging contrast and quality in this work. The results will contribute to the further advancement of the microsphere-based optical nanoscopy/microscopy techniques and facilitate their applications in nanotechnology, life sciences, etc.

\section*{Results and Discussion}

Fig.1 (a) illustrates the schematic drawing of patchy microsphere fabrication by glancing angle deposition (GAD) method~\cite{08PawarAB}. BaTiO$_3$ (BTG, 27 - 35 $\mu$m diameter, Microspheres-Nanospheres, USA) were self-assembled into monolayers by drop-casting of a small amount of BTG powders on a glass slide followed by using water to compact them together. The microsphere arrays were then coated with 100 nm thick Ag film by physical vapor deposition (PVD) (1 \AA/s) at a constant angle ($\alpha$ = 60$^\circ$). After deposition, a stream of deionized water was used to transfer the microspheres from the glass slide to the observation sample. We call the fabricated patchy BTG microsphere as p-BTG particle. 

The p-BTG particles were observed with a commercial reflected light microscope (Axio AX10, Carl Zeiss) for super-resolution imaging [Fig.1 (b)]. The microspheres collected near-field nanoscale information from the Blu-ray Disk (BD) surface and generated a magnified real image above the sample, which was then captured by the objective lens. This is confirmed by the imaging plane that is above the substrate, in contrast to virtual imaging whose image plane is down into the substrate. The entire imaging process was performed in air. A 20 $\times$ objective  (NA = 0.4, EC EPIPLAN, Carl Zeiss) were used for imaging with the p-BTG particles. The system was illuminated by a white light source (HAL 100, Carl Zeiss). All the experimental results were recorded using a high-speed scientific complementary metal-oxide-semiconductor (CMOS) camera (DFC295, Leica).

Fig.1 (c) shows the scanning electron microscopic (SEM) image of the top surface of the BD sample in this study. It has a strip pattern with 300 nm periodicity, including 200 nm track width and 100 nm gap between two adjacent tracks. The pattern of the BD cannot be resolved by conventional OM method with diffraction-limited resolution of $\lambda$/2NA = 550/0.8 = 687.5 nm [Fig.1 (d)]. The p-BTG microspheres after depositing 100 nm Ag by GAD method ($\alpha$ = 60$^\circ$) are shown in Fig.1 (e). There are some elliptical shadows on the left side of the microspheres [white arrow, Fig.1 (e)], because the microspheres blocked the transportation of Ag vapor from the source to the substrate during deposition. The corresponding SEM image of p-BTG arrays also confirms the presence of Ag patches on the microspheres [Fig.1 (f)]. Fig.1 (g) shows a p-BTG on a BD sample, in which both the Ag patch and the strip pattern can be observed.

In this study, the imaging performance of BTG and p-BTG microlenses in air were compared with each other. As shown in Fig.2 (a), the BTG in air formed a magnified, low-contrast, real image of the strip pattern above the BD. The gap between two neighboring tracks was 2.5 $\mu$m at the clearest image position, corresponding to a magnification factor (M) $\sim$ 8.3 $\times$. Here, the super-resolution image formed by BTG has a poor quality, with very low imaging contrast, which is not sufficient for most of the practical applications. On the contrast, as shown in Fig.2 (b), the new p-BTG microspheres generate significantly improved super-resolution images, both in quality and contrast. The imaging contrast has been boosted by a factor of $\sim$ 6.5 $\times$, as shown in Fig.2 (c) by retrieved intensity profile along the white dash lines in Fig. 2 (a) and (b). The measured magnification factor for p-BTG lens is about 3.9 $\times$, which is smaller than that of unpatched particles (M = 8.3 $\times$). This is caused by different focusing characteristics of BTG and p-BTG particles which will be discussed below.

Interestingly, as shown in Fig.2 (d), we observed two patchy textured pattern in each p-BTG particle, but only one side of the microspheres was coated with Ag films. The two paired patterns have a rotation angle of 180$^\circ$ between them around the center of the microsphere. This phenomenon could be attributed to the internal reflection occurring inside the high-index microspheres, in which case the Ag film prevents part of the incident light from entering the microsphere and causes a shadow with the same shape on the other side of the microsphere after multiple internal reflections.

In another imaging test, silica particles with 230 nm mean diameter (Nanorainbow Biotechnology, China) were self- assembled into arrays [Fig.2 (e)]~\cite{20FangC} and observed through BTG and p-BTG particles in air. The silica particle arrays were coated with 20 nm Ag before observation to enhance their reflectivity. As shown in Figure 2 (f), (g), the p-BTG particle again demonstrates a better imaging performance over the BTG particle when imaging a sub-diffraction-limited nanoparticle array.

In our experiments, we also found that the p-BTG particles deposited on the sample surface may have different appearance: bright, dark, or textured, and different imaging performance, depending on the positions of Ag coatings. As shown in Fig.3 (a), when the Ag film is at the bottom of the microsphere, it is like a concave mirror that reflects the incident light backwards in a convergent way, so that more light can be collected by the objective, leading to a bright appearance. On the contrary, the p-BTG looks dark when the Ag film is on top of the microsphere. As shown in Fig.3 (b), the Ag film acts like a convex mirror that reflects the incident beam divergently at a large angle, so that most of the reflected light cannot be captured by the objective. The p-BTG lens shows a textured appearance and form a magnified real image when the Ag film is on the side of the microsphere [Fig.3 (c)], in which case the Ag film acts as an aperture stop, enhancing the contrast of the image and forming a photonic hook inside the microsphere.

To understand the main focusing properties of the p-BTG microsphere lens, computational modeling was performed using the two-dimensional (2D) finite-difference-time-domain (FDTD) method (3D sphere modelling is not possible due to limited computing resource). As shown in Fig.4 (a), (b), cylinders (D = 35 $\mu$m, n$_1$ = 1.9) were created for FDTD simulation. The background medium was set to air (n$_2$ = 1). A plane light ($\lambda$ = 550 nm) propagating in the Y direction forms a photonic jet inside the cylinder [Fig.4 (a)]. Fig.4 (b) is the FDTD simulation result of the intensity field distribution in the vicinity of a cylinder partially covered with a 100 nm thick Ag film (90$^\circ$ central angle, $\theta$ = 10$^\circ$). We can see the formation of a ‘photonic hook’, with the light path off-centered and curved due to the asymmetry property of the incident beam caused by the Ag coating. In terms of imaging, the oblique near-field illumination can help the lens to capture higher orders of diffraction from the sample~\cite{18SanchezC}, which turned out to be very beneficial in boosting the imaging quality and the imaging contrast in microsphere-based super-resolution imaging. This enhancement mechanism can play an important role in developing more advanced and reliable microsphere super-resolution imaging systems.

From Fig. 4 (b) we can also see the angled ‘photonic hook’ beam leads to a larger object-to-focus distance (O). Since magnification M = I/O, where I is image plane position and O the object plane position, increasing O will lead to decreased M which can explain why the p-BTG lens produces smaller magnification factor as in experiments. 

To illustrate the position effect of Ag film on BTG particle focusing, we varied the $\theta$ angle in Fig. 4(b) from 0$^\circ$ to 80$^\circ$, while keeping the film coating opening angle $\beta$ = 90$^\circ$. As shown in Fig.4 (c), the photonic hook phenomenon is maximized at $\theta$ = 0, which gradually decreases as $\theta$ increases up to 60$^\circ$. After this angle, the focusing doesn’t show a curved hook focusing effect. In our experiments, the p-BTG particle with $\theta$ = 10$^\circ$ - 45$^\circ$ degree is recommended for overall best performance which is a balanced choice between magnification factor and imaging contrast enhancement.

\section*{Conclusion}

In conclusion, BTG microspheres with patchy coating on their surface can provide a new strategy for improving the quality of super-resolution images obtained with high-index microspheres in air. Due to the formation of photonic hook illumination condition, the super-resolution imaging contrast can be improved by more than 6-fold which significantly boost the overall imaging quality. This method enables achieving high-quality super-resolution imaging without the use of immersion liquid, such as water or oil, opening a new path to developing more advanced and reliable nano-imaging systems based on engineered microsphere lenses.  

\section*{Funding}
Postgraduate Research \& Practice Innovation Program of Jiangsu Province No.1812000024501

\section*{Disclosures} The authors declare no conflicts of interest.

\section*{acknowledge} The support of Fonds europ\'een de d\'eveloppement r\'egional (FEDER) and the Walloon region under the Operational Program “Wallonia-2020.EU” (project CLEARPOWER) is gratefully acknowledged.

\printbibliography

\newpage

\begin{figure}[hbt ]
\centering
  \includegraphics[width=0.8\linewidth]{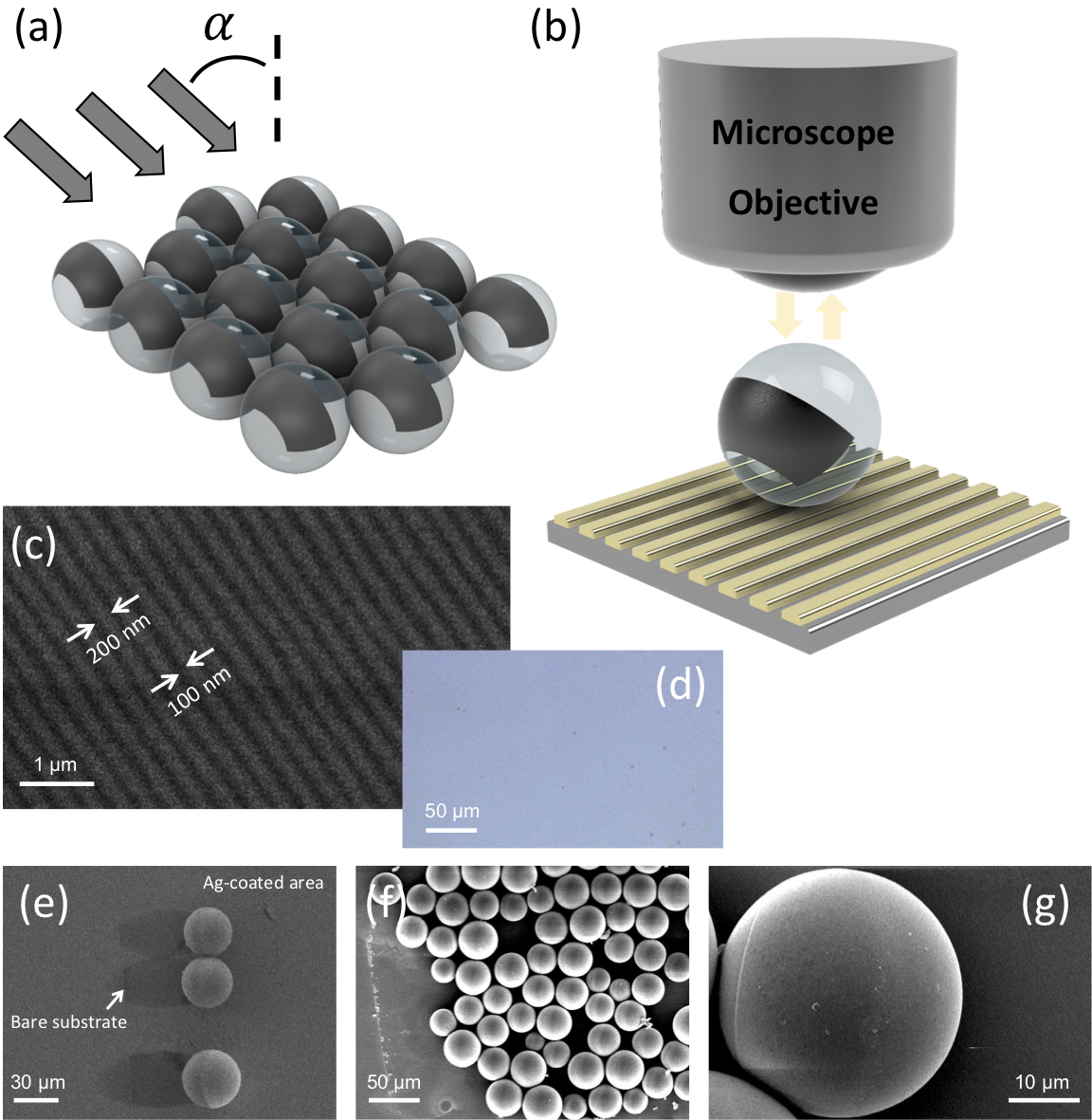}
  \caption{(a) Fabrication of patchy BTG (p-BTG) particle by glancing angle deposition method; (b) Experimental setup of super-resolution imaging with p-BTG lens; (c) SEM image and (d) OM image of BD substrate (lines not resolved); (e), (f) SEM images of p-BTG particle after Ag deposition; (g) SEM image of p-BTG particle transferred on BD substrate.}
  \label{Fig.1}
\end{figure}

\begin{figure}[hbt]
  \centering
  \includegraphics[width=0.85\linewidth]{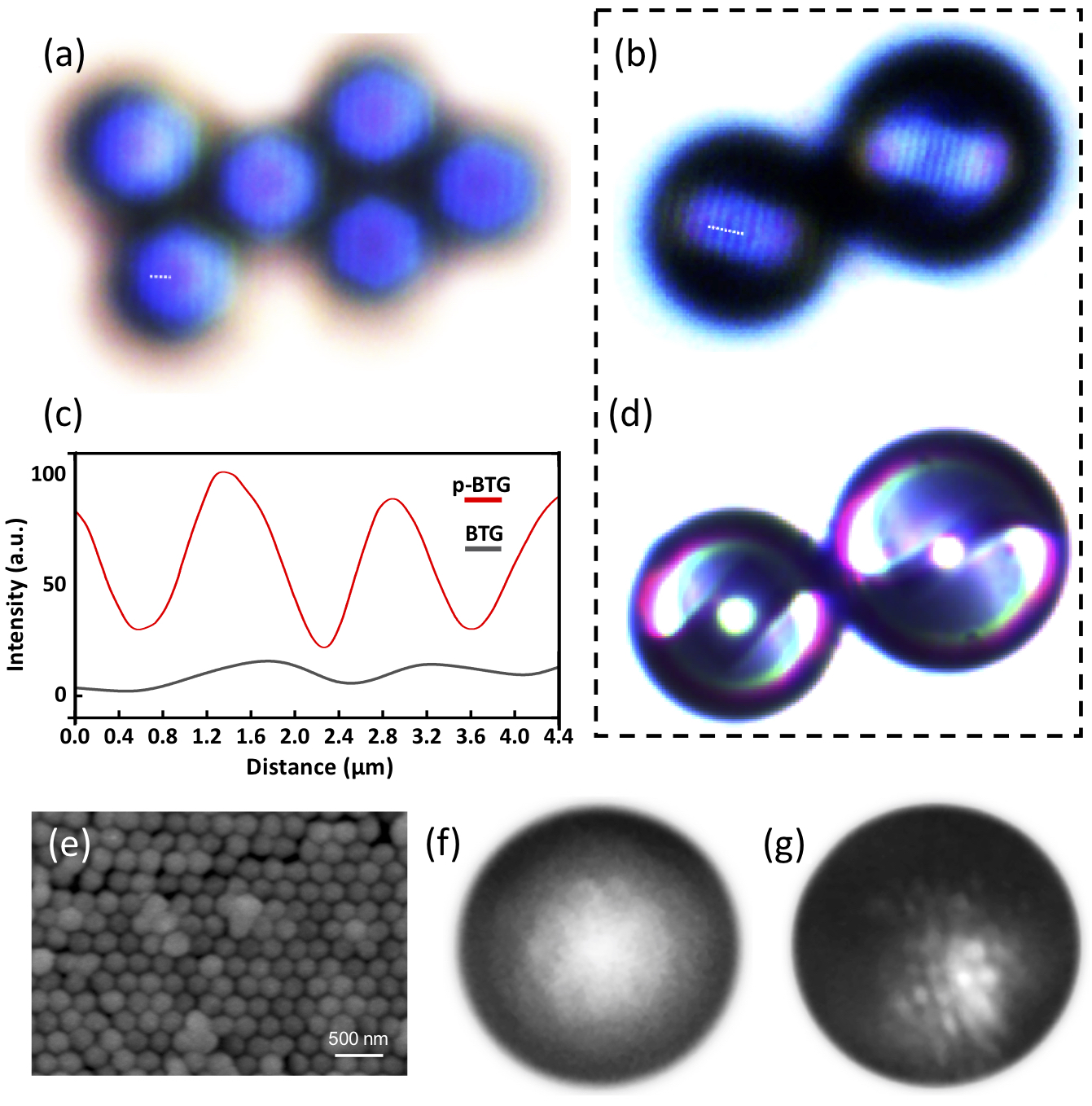}
  \caption{(a), (b) OM images of the pattern on BD surface observed through (a) pristine BTG and (b) p-BTG with the 20 $\times$ objective; (c) Optical intensity profiles across the white dash lines; (d) OM image of the corresponding p-BTG; (e) SEM image of the 230 nm-diameter SiO$_2$ particle arrays; (f), (g) The SiO$_2$ particle arrays observed through (f) pristine BTG and (g) p-BTG microspheres.}
\label{Fig.2}
\end{figure}

\begin{figure}[hbt]
\centering
 \includegraphics[width=0.9\linewidth]{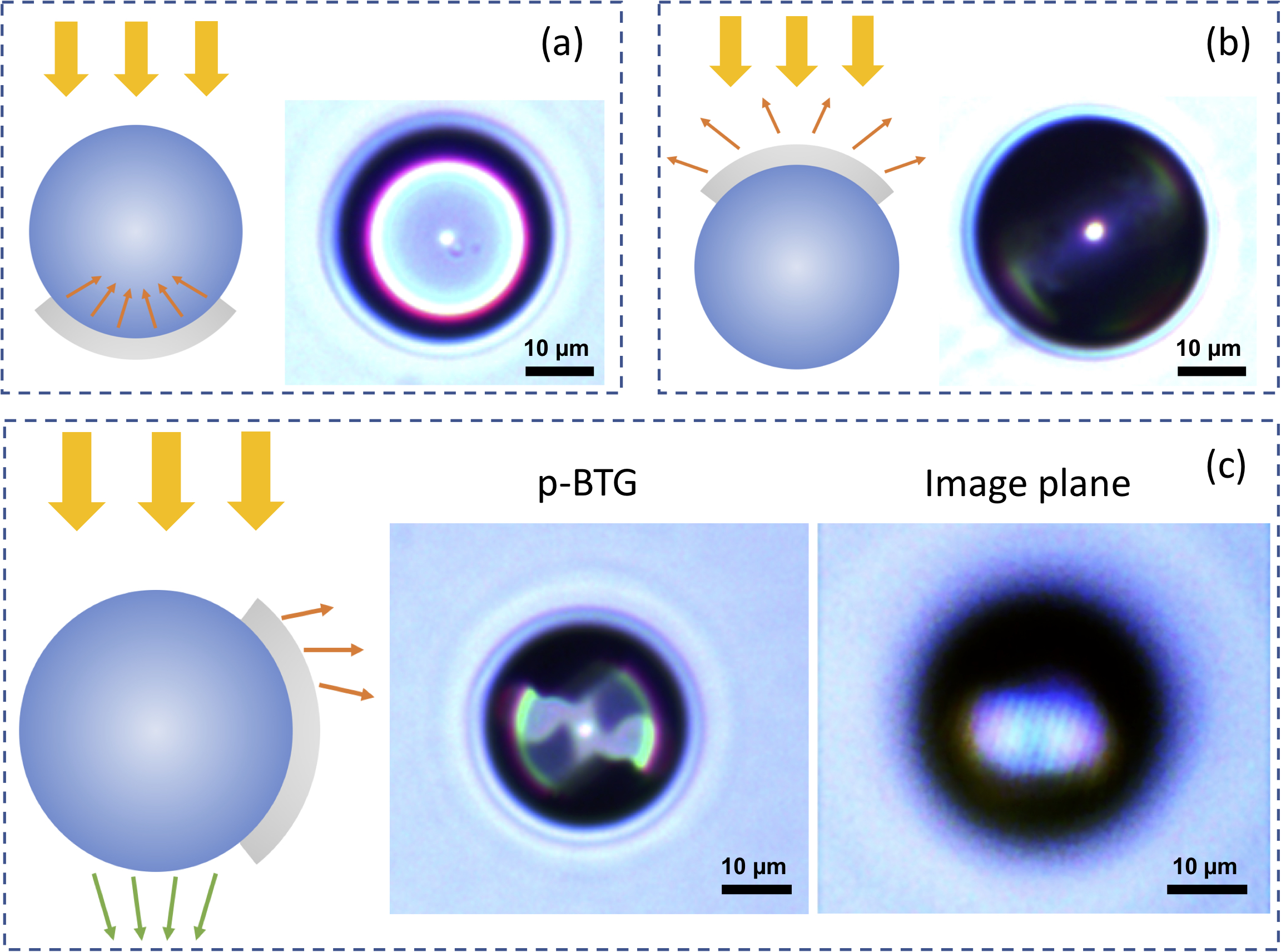}
 \caption{OM images of p-BTG particles when the Ag film is (a) at bottom of the microsphere, (b) on top of the microsphere and (c) on the side of the microsphere.}
 \label{Fig.3}
\end{figure}

\begin{figure}[hbt]
\centering
  \includegraphics[width=0.9\linewidth]{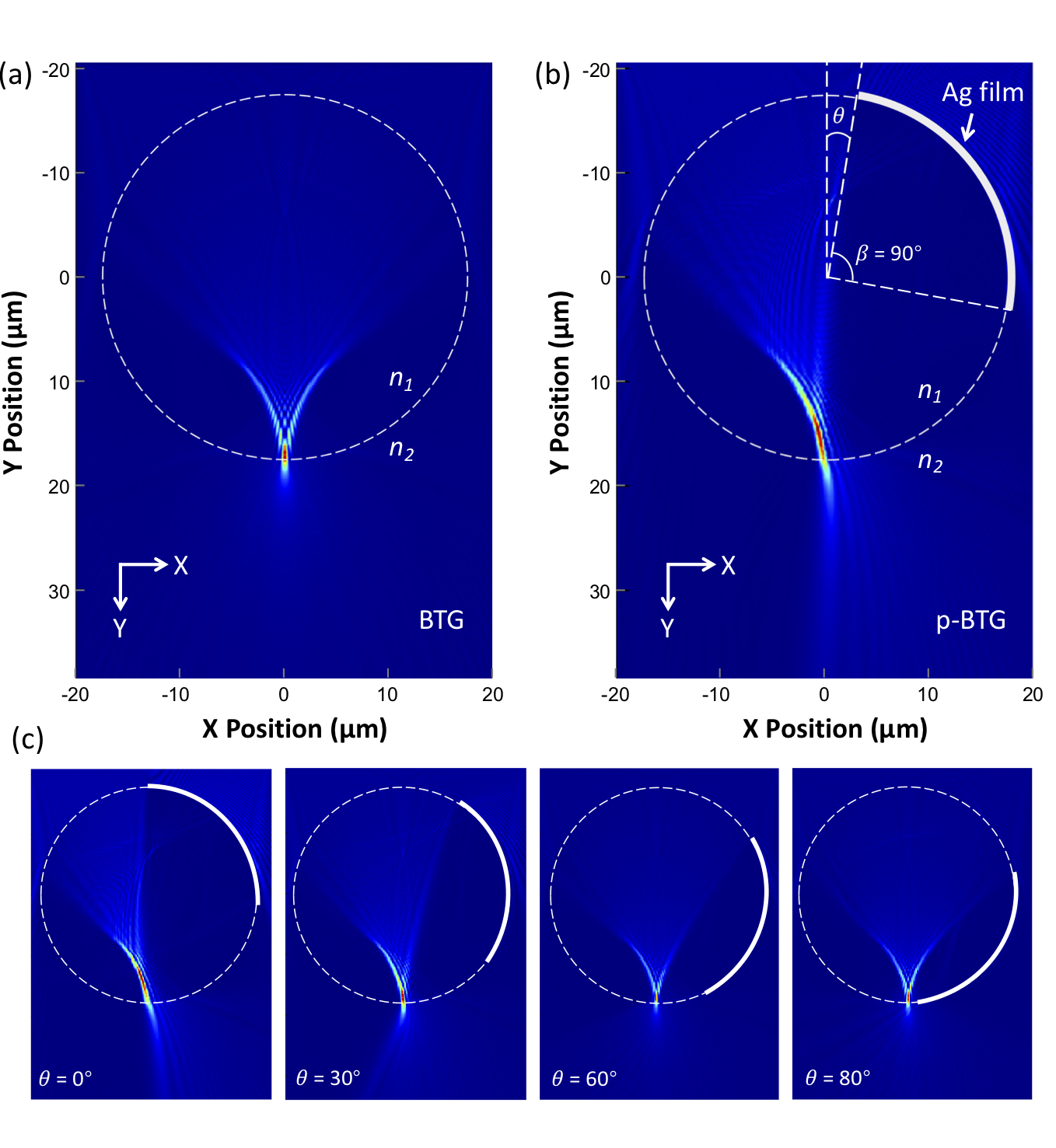}
  \caption{(a), (b) FDTD-simulated light field of (a) the pristine BTG and (b) the p-BTG; (c) The influence of the position of Ag films on the focusing of the p-BTG.}
  \label{Fig.4}
\end{figure}

\end{document}